\documentclass[useAMS,usenatbib]{mn2e}

\usepackage{natbib, aas_macros,color}
\usepackage{ulem}
\usepackage[dvips]{graphicx}
\usepackage{wrapfig}
\graphicspath{{./figs/}}
\usepackage{color}
\usepackage{amsmath}
\usepackage{amssymb}

\newcommand{\Msun}{M_{\odot}}

\newcommand{\Zsun}{Z_{\odot}}
\newcommand{\fesc}{f_{\rm esc}}

\newcommand{\nion}{\dot{n}_{\rm Ion}^{\gamma}}

\newcommand{\lya}{\rm {Ly{\alpha}}}

\newcommand{\Msunyr}{\rm M_{\odot}~ yr^{-1}}

\newcommand{\Tb}{\delta T_{\rm b}}
\newcommand{\Ts}{T_{\rm S}}
\newcommand{\Tcmb}{T_{\rm CMB}}

\newcommand{\Qhii}{Q_{\rm HII}}
\newcommand{\Tgas}{T_{\rm gas}}
\newcommand{\xa}{x_{\rm \alpha}}
\newcommand{\Ja}{J_{\rm \alpha}}
\newcommand{\ergs}{{\rm erg~s^{-1}}}
\newcommand{\Thi}{T_{\rm HI}}
\newcommand{\taue}{\tau_{\rm e}}
\newcommand{\tb}{\delta T_{\rm b}}
\newcommand{\xhii}{X_{\rm HII}}

\title[Can the 21 cm signal probe Pop III and II SF?]
{Can the 21 cm signal probe Population III and II star formation?}

\author[Yajima et al.]
{Hidenobu Yajima$^{1, 2}$\thanks{E-mail: yajima@roe.ac.uk (HY)} and Sadegh Khochfar$^{1}$
\\
$^{1}$ SUPA\thanks{Scottish Universities Physics Alliance}, 
Institute for Astronomy, University of Edinburgh, Royal Observatory, Edinburgh, EH9 3HJ, UK\\
$^{2}$ Department of Earth \& Space Science, Graduate School of Science, Osaka University, 1-1 Machikaneyama, 
   Toyonaka, Osaka 560-0043, Japan\\
}
 
\begin{document}

\date{Accepted ?; Received ??; in original form ???}

\pagerange{\pageref{firstpage}--\pageref{lastpage}} \pubyear{2008}

\maketitle

\label{firstpage}

%
%
\begin{abstract}
Using varying models for the star formation rate (SFR) of Population (Pop) III and II stars at $z>6$ we derive the expected redshift history of the global 21 cm signal from the inter-galactic medium (IGM).
To recover the observed Thomson scattering optical depth of the cosmic microwave background (CMB) requires SFRs at the level of $\sim 10^{-3}~\rm M_{\rm \odot}~yr^{-1}~Mpc^{-3}$ at $z \sim 15$ from Pop III stars, or  $\sim 10^{-1}~\rm M_{\rm \odot}~yr^{-1}~Mpc^{-3}$ at $z \sim 7$ from Pop II stars. 
In the case the SFR is dominated by Pop III stars, the IGM quickly heats above the CMB at $z \gtrsim 12$ due to heating from supernovae. 
In addition, $\lya$ photons from haloes hosting Pop III stars couple the spin temperature to that of the gas, resulting in a deep absorption signal. 
If the SFR is dominated by Pop II stars, the IGM slowly heats and exceeds the CMB temperature 
at $z \sim 10$.
However, the larger and varying fraction of  Pop III stars is able to  break this degeneracy.   
We find that the impact of the initial mass function (IMF) of Pop III stars on the 21 cm signal results in an earlier change to  a positive signal  if the IMF slope is $\sim -1.2$. Measuring the 21 cm signal at  $z \gtrsim 10$ with next generation radio telescopes such as  the Square Kilometre Array  will be able to investigate the contribution from Pop III and Pop II stars to the global star formation rate.
\end{abstract}

%
%

%
%
\section{Introduction}
The star formation history (SFH) of galaxies and the physical state of the intergalactic medium (IGM) at $z \geq 6$ is of key importance to understand the transition of the Universe from a neutral to highly ionized state and the sources responsible for it.
The observational situation has made significant progress over the recent years. The Thomson scattering optical depth of the cosmic microwave background (CMB) suggest that the IGM was re-ionized by $z \sim 11$ \citep{Komatsu11, Planck13}. In addition,  $\lya$ emitters  allow to estimate the degree to which the IGM is re-ionized at $ z \sim 7$ \citep{Ota08, Kashikawa06, Kashikawa11}.
As for the galaxy population, the {\it Hubble Space Telescope} (HST) with Wide Field Camera 3 (WFC3) allowed to derive stellar masses and star formation rates for galaxies at $z > 6$ \citep[e.g.,][]{McLure11, Bouwens12}. 
These are complemented by afterglow measurements of $\gamma$-ray burst, which suggest similar cosmic star formation histories at  $z \sim 6-8$ \citep{Yuksel08, Kistler09, Kistler13}. 
Despite these successes, the physical interpretation of the observations is still partly degenerate as to the detailed redshift evolution of the SFH and  physical state of the IGM at $ z \geq 6$.
This is mainly because the CMB observations are interpreted assuming instantaneous reionization,  and observations of the galaxy population only probe the bright-end $( > L_*)$ of the luminosity function at high redshifts securely \citep{Ono12, Shibuya12, Finkelstein13}.  
 Moreover, the SFH of Population (Pop) III stars and their impact on the IGM is very complex. 
Despite recent state-of-the-art simulations of Pop III stars formation \citep[e.g.][]{Yoshida08, Turk09, Clark11, Stacy12, Wise12b, Ahn12, Umemura12, Susa13, Hirano14, Stacy14}, understanding of their formation over a wide range of physical conditions and redshift range ($z = 6-30$) is still lacking.
One promising way to resolve these issues is the observation of the 21 cm line from the IGM at $z > 6$ by the LOw Frequency ARray \citep[LOFAR;][]{Harker10}, the Murchison Widefield Array \citep[MWA;][]{Lonsdale09}, and the Square Kilometre Array \citep[SKA;][]{Dewdney09}.  
For example, LOFAR covers $z \lesssim 11$ with the angular resolution of $\sim 3"$,
and SKA will cover more wide range surveys at $z \lesssim 19$ with a higher resolution of $\sim 1"$.

The 21 cm emission is sensitive to the ionization state of hydrogen, gas temperature and the $\lya$ radiation field,
 which are controlled by the first stars, galaxies, and quasars (QSOs) \citep{Yajima13b}.
Previous works has addressed the detailed structures of 21 cm emission around 
Pop III stars \citep{Chen04, Chen08, Tokutani09, Yajima13b},
galaxies \citep{McQuinn06, Mellema06b, Kuhlen06, Wyithe07, Semelin07, Baek09, Mesinger11, Iliev12}, and 
QSOs \citep{Wyithe05, Geil08, Alvarez10, Datta12, Yajima13b}.
Their work  suggests that future observations are able to resolve the 21 cm structure around giant H{\sc ii} region made by massive galaxies like QSO hosts, 
while it will be difficult to detect individual sources around typical galaxies or Pop III stars \citep{Yajima13b}.
The distribution of giant H{\sc ii} bubbles as probed by 21 cm observation will thus give us valuable information about structure formation and cosmology \citep[see also,][]{Chongchitnan12}.

An alternative approach is to focus on the redshift history of the global 21 cm emission, focusing on the spatially averaged value instead \citep{Furlanetto06a, Pritchard08, Mirocha13}.
Instead of the detailed structure around sources, the global signal reflects the statistical nature of stars and the IGM, 
e.g., the global star formation history of Pop II \& III stars, the mean degree of ionization and temperature of the IGM, and the mean intensity of $\lya$ radiation.  
Previous studies investigated the redshift evolution of the differential brightness temperature ($\tb$) with different free parameters for the  heating rate and SFR \citep{Pritchard08}. 
In this work, we revisit earlier attempts by taking into account recent progresses on cosmic reionization and the star formation history of galaxies.  We construct models that match the latest observations, and investigate effects of Pop II \& III stars on the redshift history of the global 21 cm emission. 

Our paper is organized as follows.
We describe our model in Section~\ref{sec:model}. 
In Section~\ref{sec:result}, we show the redshfit evolution of the ionization degree, temperature, and 21 cm signal of the IGM. 
In Section~\ref{sec:discussion}, we investigate the effects of black holes and initial mass function of Pop III stars on the 21 cm signal. 
Finally, in Section~\ref{sec:summary}, we summarise our main conclusions.

%
%
\section{Model}
\label{sec:model}
In the following sections we lay out our model assumptions. Throughout this paper we use the cosmological parameters $\Omega_{\rm M}=0.26, \Omega_{\rm \Lambda} = 0.74$ and $h = 0.72$ as inferred by the WMAP 7-year data \citep{Komatsu11}, 
which are consistent with the results of the Planck mission \citep{Planck13}, and do not change the results of our study significantly. 

\subsection{Star formation history}

We start by constructing  models of the Pop II and III SFH. 
A common fit to the Pop II SFRD is given in \citet{Bouwens11} as:
\begin{equation}
\dot{M}_{\star}^{\rm PopII}= (a + bz)h / \left[ 1 + (z/c)^{d} \right]~\rm \Msun \; yr^{-1} \; Mpc^{-3}
\end{equation}
where $(a,b,c,d)$ are fitting parameters, with best fitting values  $a = 0.017$, $b=0.13$, $c=3.3$ and $d=5.3$. 
In the following we will refer to this as our fiducial model.
There is some degree of discussion as to the appropriate values for the free parameters. 
For example, $d=7$ fits the SFRD from the galaxy population reported in  \citet{Oesch13}, 
while $c=4.5$ matches GRBs observations by \citet{Yuksel08, Kistler09, Kistler13}.
The latter discrepancy has been argued is a result of GRBs probing star formation rates in galaxies below the detection limit of surveys. Integrating the observed luminosity function of galaxies under the assumption of a faint-end slope    $\alpha \sim - 1.7$  down to $L=0$ brings both estimates in agreement.  
We, however note that the faint-end slope is highly uncertain within  the current detection limits of galaxy observations \citep[e.g.,][]{Khochfar07, McLure11}   and the simple fit to the  luminosity function  may be broken at low luminosity due to the strong suppression of star formation in low-mass galaxies by radiative and supernovae feedback \citep{Wise12a, Hasegawa13}.
In addition, the sample of GRBs at $z > 6$ is still very small ($\sim$ a few). 
Therefore, considering the uncertainties, we investigate two alternative star formation models, a high star formation model (Pop IIb) with $d = 7.0$ and a low one (Pop IIc\&d) with $c=4.5$. 
Moreover, recent galaxy observations suggested that SFRD steeply decreases with increasing redshift at $z > 8$ \citep[e.g.,][]{Oesch13}.
Theoretical modeling based on the collapse fraction of haloes showed SFRD falls exponentially at high-redshift \citep[e.g.,][]{Pritchard10, Mirocha13}.
Therefore, as an additional model, we consider the exponential decay model (PopIIe) at $z>8$ with
${\rm SFRD} = b\times{\rm exp}(-a\times z)$ where $a$ and $b$ are fitting parameters. 
The PopIIe model follows PopIIa at $z \le 8$. The fitting parameters are chosen to match the observational data by \citet{Oesch13, Oesch14},
and they are $a=1.6$ and $b=7.0$ respectively.
A summary of the models is listed  in Table~\ref{table:model}. 
The number of $\lya$ and ionizing photons emitted from the source stars is computed based on the theoretical spectral energy 
distribution (SED) given by the population synthesis code P\'{E}GASE v2.0 \citep{Fioc97} with a metallicity of $Z = 10^{-2} \Zsun$ \citep[e.g.,][]{Wise12a} and Salpeter IMF.

\begin{figure}
\includegraphics[scale=0.7]{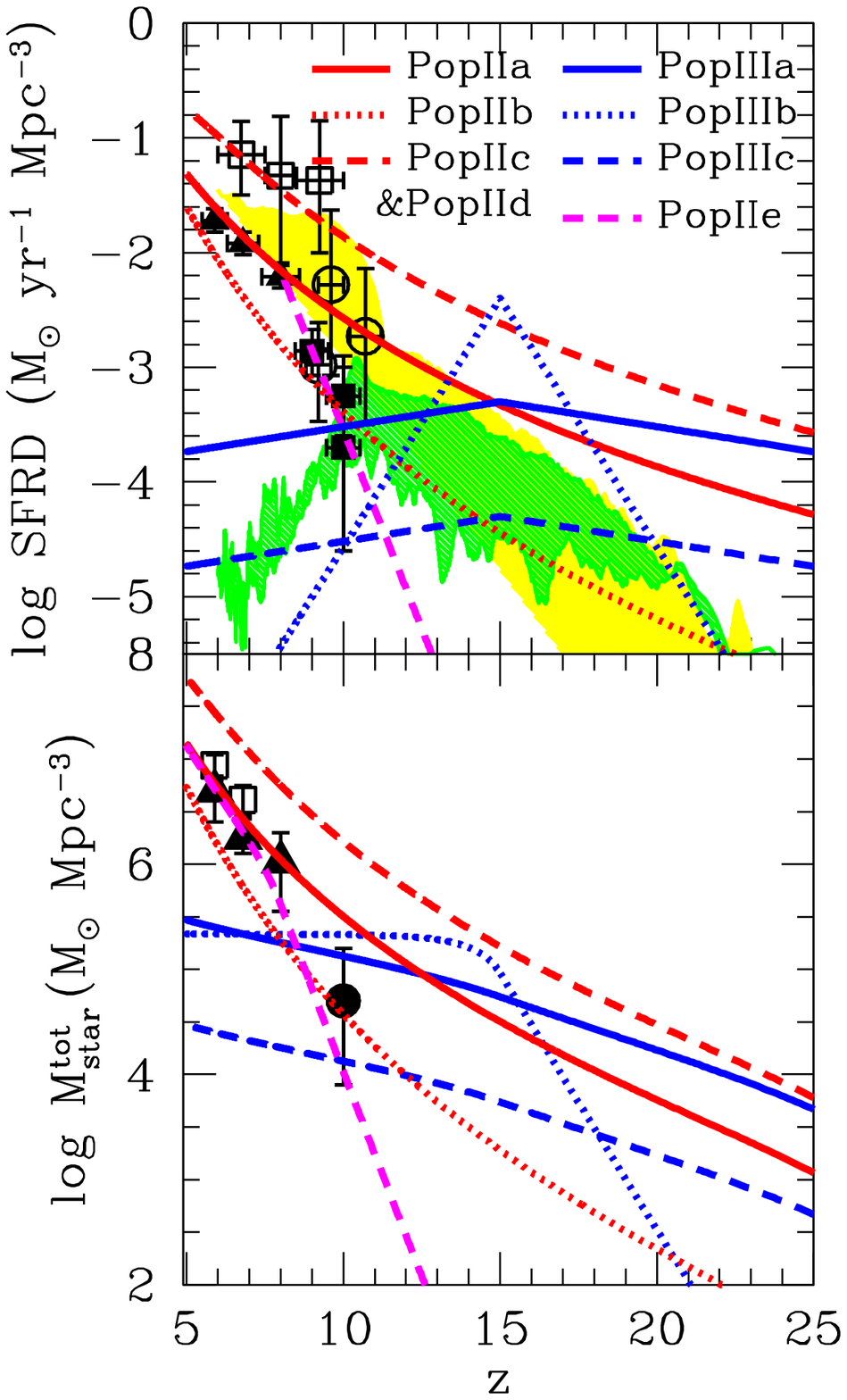}
\caption{
{\it Upper panel}: Redshift evolution of cosmic star formation rate density. 
Different lines represent different star formation models shown in Table~\ref{table:model} and Table~\ref{table:popiiimodel}.
Symbols show the  observational data. Filled triangles are from \citet{Bouwens07, Bouwens12}, and filled squares are from \citet{Oesch13, Oesch14}. 
Open circles are from CRASH cluster detections \citep{Bouwens12b, Zheng12, Coe13}.
Open squares are derived from gamma-ray bursts \citep{Kistler13}.
Green and yellow shade regions represents the star formation rate densities of Pop III and II stars respectively, calculated in the cosmological simulations by \citet{Johnson13}. The range of y-axis shows the models with or without the Lyman-Werner UV background which can suppress formation of hydrogen molecules. 
{\it Lower panel}: Cumulative stellar mass density of Pop III and II stars with different models.
Symbols represent the observational data from \citet[][open squares]{Gonzalez11},
\citet[][filled triangles]{Stark13}, and \citet[][filled circle]{Oesch14}.
}
\label{fig:sfr}
\end{figure}

On the other hand, for the Pop III stars, the $\lya$ and ionizing photon emissivities are not sensitive to the IMF, it is roughly proportional to total stellar mass alone,
because the effective temperature is almost constant at $T \sim 10^{5}~\rm K$ and the luminosity scales close to linearly with the mass \citep{Bromm01a}.
Hence we use the ionizing photon emissivity of $120~\Msun$ stars derived by \citet{Schaerer02} and estimate the total ionizing photon emissivity per solar mass by dividing by $120~\Msun$, and assume the effective temperature of $T=10^{5}~\rm K$ to estimate $\lya$ photons from stellar continuum radiation. 
In addition, there are no strong observational constraints on the PopIII SFR.
In the case of Pop III stars, their formation rate  is reduced due to metal enrichment by supernovae \citep[e.g.,][]{Omukai08, Maio11, Wise12a, Johnson13}. 
As a result, the SFRD of Pop III stars might have a peak at a specific redshift.
Hence, we use the simple exponential decay model with a peak at the typical Pop III star forming redshift $z = 15$ \citep[e.g.,][]{Yoshida08, Johnson13} as follows,
\begin{equation}
\dot{M}_{\star}^{\rm PopIII} = A \times{\rm exp} (- |z - 15|/ B) ~\rm \Msun \; yr^{-1} \; Mpc^{-3}
\end{equation}
where $A$ and $B$ are again fitting parameters. The parameter ``$A$'' is the amplitude factor, and the ``$B$'' controls the steepness of the redshift dependence. 
Even with state-of-the-art simulations of Pop III stars, there is a large room in the mean values and redshift evolution, because of the difficulties to resolve the initial mass function, the metal/dust enrichment  and radiative/supernovae feedback. 
Recent radiative hydrodynamics simulations show a weak redshift evolution with ${\rm SFRD} \sim 10^{-4} - 10^{-5}~\Msunyr$ \citep{Wise12a}.
Therefore, we consider slow redshift evolution models ($B=10.0$) with a high SFRD (PopIIIa: $A=10^{-3}$) and a lower one (PopIIIc: $A=10^{-5}$).
In addition, we construct a rapid redshift evolution model (PopIIIb) with $A=10^{-2}$ and $B=1.0$.
These star formation models are shown in Figure~\ref{fig:sfr}. We see the SFHs of Pop II stars satisfy recent observations.
In addition, recent cosmological simulations by \citet{Johnson13} including supernova feedback predict star formation rates similar to those assumed by our models.
The bottom panel of the Figure~\ref{fig:sfr} shows the cumulative stellar mass as a function of redshift. Current observation are still highly uncertain at $z \gtrsim 6$. 
Note that we here estimate the stellar mass by integrating the SFRD.
Our PopIIa or PopIIb models roughly recover the observed stellar mass density (SMD) at $z \le 8$, while they are above the observed ones at $z \sim 10$.
However, the SMD at $z \sim 10$ was estimated from only four bright galaxies. 
Therefore, there is a large uncertainty in the contribution from low-mass galaxies. 
If the faint-end slope is steeper, the SMD can increase significantly.  
The PopIIe follows the observed SMD at $z \sim 5 - 10$. 
In the case of massive Pop III stars, their life times are short and they end as either  Type-II supernovae or direct collapse to black holes \citep{Heger02}. As a result, the Pop III cumulative mass is an overestimate of the actual stellar mass.

The SFRD in model PopIIa is higher than that of PopIIIb and PopIIIc at $z \gtrsim 18$ in contrast to expectations from models. 
However, the contribution of PopII stars in this model towards cosmic reionization and the 21 cm signal at $z \gtrsim 18$ is negligible and does not impact our conclusions.
For the PopIIa+PopIIIb model, we turn off the formation of Pop II stars when the SFRD of Pop II stars is higher than that of Pop III stars at $z \gtrsim 18$.

\subsection{Ionization history}
Based on the above models for the SFRD, 
we derive cosmic reionization histories. 
Some fraction of ionizing photons from stars can escape from halos and ionize the surrounding IGM \citep{Hasegawa13, Paardekooper13}.
Recent theoretical work shows the escape fraction is $\sim 0.2$ for typical galaxies \citep{Yajima09, Yajima11, Yajima12d, Razoumov10},
and $\sim 0.5$ for dwarf galaxies or Pop III stars \citep{Abel07, Whalen04, Kitayama04, Paardekooper13}.
Here, we assume $\fesc = 0.5$ for Pop III stars and $\fesc = 0.2$ for Pop II stars for our fiducial models, and $\fesc = 0.5$ for one additional model of the Pop II SFH (PopIId).
Using these models the  redshift evolutions of volume fraction of H{\sc ii} region is calculated as  \citep{Barkana01}: 
\begin{equation}
\frac{d\Qhii}{dt} =  \frac{1}{n_{\rm H}^{0}} \nion - \alpha_{\rm B} C (1+z)^{3} n_{\rm H}^{0} \Qhii
\end{equation}
where $\Qhii$ is the volume fraction of H{\sc ii}, $n_{\rm H}^{0}$ is the present-day hydrogen number density ($\sim 1.9 \times 10^{-7}~\rm cm^{-3}$),
$\nion$ is a number density of the escaped ionizing photon, $\alpha_{\rm B}$ is the case-B recombination rate, and $C$ is a clumpiness factor of IGM. 
We assume the recombination rate at  $T=10^{4}~\rm K$ ($\alpha_{\rm B} = 2.59\times10^{-13}~\rm cm^{3}\;s^{-1}$) 
and $C=3$ which is suggested by numerical simulations \citep{Pawlik09b, Jeon14}.
Free electrons produced by cosmic reionization contributes to the Thomson scattering optical depth ($\tau_{\rm e}$) in CMB observations which is estimated by
\begin{equation}
\tau_{\rm e} = \int_{0}^{\infty} \sigma_{\rm T} n_{\rm e}(z)  c \left| \frac{dt}{dz} \right| dz,
\end{equation}
where
\begin{equation}
\frac{dt}{dz} = \frac{\rm 1}{(1+z)H_{0}\sqrt{\Omega_{\rm M}(1+z)^{3}+\Omega_{\rm \Lambda}}}.
\end{equation}
In this work, we assume the single ionization rate of helium is the same as the one for hydrogen 
at $z \ge 3$, and the double ionization takes place at $z < 3$ \citep[e.g.,][]{Wyithe10, Inoue13}.
Recent simulations show that indeed the fraction of He{\sc ii} is close to the H{\sc ii} fraction at the high redshift, although the ionization fraction of helium is slightly lower than the one for hydrogen \citep{Ciardi12}.

\subsection{Thermal evolution}

Before $z \sim 150$, the temperature of the gas is given by its coupling to the CMB  via Compton scattering \citep{Peebles93}. 
Thereafter, gas starts to decouple from the CMB and the temperature decreases by adiabatic expansion $(1+z)^{2}$, while that of the CMB does as  $(1+z)$. However, as structure grows, additional heating sources can form. 
In particular, X-ray photons can travel for long distance and partially ionize the IGM, and
 become important sources for the global 21 cm emission \citep{Furlanetto06a, Mirocha13}.
 To account for this  we include the X-ray emission from supernovae as estimated in   \citet{Furlanetto06a},
\begin{equation}
L_{\rm X}^{\rm SN} = 1.6 \times 10^{40} f_{\rm e} \left(
\frac{\epsilon_{\rm e}}{0.05} \frac{\nu_{\rm SN}}{0.01 \rm \Msun^{-1}} \frac{E_{\rm SN}}{10^{51} \rm erg} \frac{\rm SFR}{1 \Msunyr}
\right) ~\ergs,
\label{eq:sn}
\end{equation}
where $f_{\rm e}$ is the deposited energy fraction from accelerated electrons in supernova shocks into X-ray energy,
$\epsilon_{\rm e}$ is the conversion fraction from total supernova energy to the acceleration of electrons, 
$\nu_{\rm SN}$ is the number of supernovae per unit mass of star formation, 
and $E_{\rm SN}$ is the total supernova energy. 
Following \citet{Furlanetto06a}, we use the values of $f_{\rm e} = 0.5$, $\epsilon_{\rm e}=0.05$
$\nu_{\rm SN}= 0.01 ~\Msun^{-1}$, $E_{\rm SN}=10^{51}~\rm erg$ for Pop II stars.
Hence, the X-ray luminosity from Pop II stars is simply given by 
$L_{\rm PopII, X}^{\rm SN} = 0.8 \times 10^{40} (\rm SFR / \Msunyr) ~\ergs$.
The typical mass of Pop III stars is likely to be much higher than that of Pop II due to the higher Jeans mass and accretion rate at their formation \citep[e.g.,][]{Bromm01b, Nakamura01, Abel02a, Yoshida08, Hosokawa11, Umemura12}. 
Very recently, \citet{Hirano14} suggested that the mass of Pop III stars significantly changes depending on the formation sites
and that the initial mass function might be flat with the mass ranging from $\sim 10$ to $\sim 500~\Msun$. 
Hence, $\nu_{\rm SN}$ should be higher than $0.01~\Msun^{-1}$ for Pop III stars.
We here assume a  flat shape with mass range $10-500~\Msun$ giving $\nu_{\rm SN} \sim 0.17$ \citep{Heger02}. 
In addition, Pop III stars of $120-260~\Msun$ end their lifes as  pair-instability supernovae with energies  $\sim 10^{53}~\ergs$.
As a result, the X-ray luminosity is estimated by $L_{\rm popIII, X}^{\rm SN} = 1.0 \times 10^{43}  (\rm SFR / \Msunyr) ~\ergs$. 
In this work, we use the flat IMF in our fiducial model, and then investigate the impact of different IMF slopes on the 21 cm signal. 
The energy fraction from  X-rays that goes into  heating the IGM ($f_{\rm X, heat}$) is estimated by 
$f_{\rm X, heat} = 0.9971 \left[ 1 - (1-x_{\rm HII})^{1.3163} \right]$ \citep{Shull85}.
Considering the X-ray heating, the temperature evolution of H{\sc i} gas is derived from \citep{Maselli03, Yajima13b}
\begin{equation}
\frac{d\Thi}{dt} = \frac{2}{3k_{\rm B}n_{\rm H}} \left[ k_{\rm B}\Thi\frac{dn_{\rm H}}{dt} + \Gamma_{\rm X, heat} \right],
\label{eq:temp}
\end{equation}
where $\Thi$ is the temperature of neutral hydrogen gas and $\Gamma_{\rm X, heat}$ is the X-ray heating rate.
The first term on the right hand side of the equation represents the cooling due to cosmic expansion.
The contribution from supernovae to cosmic reionization is not significant \citep{Johnson11}.
Hence we do not include the ionization by X-ray photons.

\subsection{$\lya$ background radiation}

With the onset of star formation the $\lya$ background field starts appearing.
$\lya$ radiation plays a key roll in determining the spin temperature via the Wouthuysen-Field effect \citep{Wouthuysen52, Field58, Hirata06}. 
There are mainly two sources of $\lya$ radiation, 
one is due to the recombination processes in H{\sc ii} regions,  another is stellar continuum radiation in the frequency range from the Lyman-limit to $\lya$. 
Stellar continuum radiation contributes to $\lya$ photons
 via the transition to the higher levels of the Lyman series ($\rm  Ly\gamma, Ly\delta..$) and cascade decay passing through from the 2p to the 1s state or the frequency shift to $\lya$ by cosmological redshifting.
Previous studies of the global 21 cm emission, only took into account continuum stellar radiation from PopII because of its dominant contribution with respect to recombinations in the IGM or haloes.   
However, in the case of Pop III stars, the $\lya$ luminosity from the recombination process is larger than that of the continuum radiation due to the high effective temperature \citep{Bromm01a} of the star.
In addition, due to the small size of H{\sc ii} regions ($\sim \rm kpc$), $\lya$ photons can be efficiently scattered at the outer H{\sc i} edge \citep{Yajima13b}.
On the other hand, during the era when Pop II stars are dominant sources of ionization, 
the typical size of HII bubble can be lager \citep{Iliev12, Wise12a}. 
Therefore, $\lya$ photons from halos may travel in the neutral IGM without scatterings due to the large velocity offset by the Hubble flow.
In this work, hence, we apply $\lya$ photons from haloes for the calculation of $\Ts$ when the SFRD of Pop III stars is higher than that of Pop II stars or $X_{\rm HII} < 0.5$.  However, note that, even at lower redshift $z \sim 7$,  recent observation suggested a large fraction of $\lya$ flux  is scattered because of surrounding residual H{\sc i} gas \citep{Kashikawa11}. 
Theoretically, \citet{Laursen11} showed that 0.8 of $\lya$ photons from galaxies at $z=6.5$ were scattered by the IGM, despite most of the IGM was ionized. 
The IGM transmission to $\lya$ photons is complicated because of the inhomogeneous HI and velocity distribution around galaxies, hence it is still controversial. 
Here, we use this simple model and will investigate detailed effects of $\lya$ photons from haloes on the IGM by numerical simulations in future work.  
For $\lya$ photons from the recombination in the IGM, the position of sources can be close to neutral IGM patches, hence some of them are scattered even when the IGM is highly ionized. 
Hence, we consider $\lya$ photons from the IGM for all cases. This may overestimate the number of $\lya$ scattering per atom, however, the effect of $\lya$ photons from the IGM is negligible because the emissivity is quite small due to the long recombination time scale of the IGM as shown in the later section. 
Thus, we estimate the $\lya$ emissivity and its relation  with the 21 cm signal as follows:
\begin{equation}
\epsilon_{\lya} = \epsilon_{\rm IGM}^{\rm rec} + \epsilon_{\rm ISM}^{\rm rec} + \epsilon_{\rm star}^{\rm cont}
\end{equation}
where
\begin{equation}
\epsilon_{\rm IGM}^{\rm rec}  = 0.68 h\nu_{\rm \alpha} n_{\rm H}^{2} (1+ f_{\rm He}) \alpha_{\rm B} Q_{\rm HII},
\end{equation}

\begin{equation}
\epsilon_{\rm ISM}^{\rm rec}  = 0.68 h\nu_{\rm \alpha} \dot{n}_{\rm Ion} (1 - f_{\rm esc}),
\end{equation}

\begin{equation}
\epsilon_{\rm star}^{\rm cont}  = f_{\rm conv} \times \int_{\rm 912\AA}^{\rm 1216\AA} \frac{L_{\nu}}{h\nu} d\nu.
\end{equation}

The $\epsilon_{\rm IGM}^{\rm rec}$,  $\epsilon_{\rm ISM}^{\rm rec}$ and  $\epsilon_{\rm star}^{\rm cont}$   are the $\lya$ emissivities from H{\sc ii} regions in the IGM,
ISM, and the stellar continuum radiation, respectively. The $\nu_{\rm \alpha}$ is the $\lya$ frequency.
The factor 0.68 is the average probability of $T=10^{4}~\rm K$ gas that a recombination process emits a $\lya$ photon via the 2p-1s state transition. The $f_{\rm conv}$ is the conversion probability from the continuum to $\lya$ line, it is $0.63$ for Pop III stars
and $0.72$ for Pop II stars \citep{Pritchard06}. These different values are due to the different slope of SEDs at the frequencies between $\lya$ and Lyman limit. 
As explained above, $\epsilon_{\rm ISM}^{\rm rec} = 0$ for SFRD(Pop II) $>$ SFRD(Pop III) or $X_{\rm HII} > 0.5$.
Note that, here $\lya$ photons from recombination in the IGM or ISM can be localized, compared to stellar continuum radiation. However, these recombinations are the dominant sources for Pop III stars, and H{\sc ii} bubbles made by Pop III stars are typically much smaller than the angular resolution of future 21 cm observation. Hence these photons can work on the global 21 cm emission signal. The mean values may differ between our simple model and coarse grained ones from detailed 21 cm structures around sources. 
We will investigate the detailed structure of 21 cm signal by using cosmological simulations in future work.
In addition, even if $\lya$ photons are scattered by the surrounding IGM,  
the propagation distance from Pop III stars can be small during the early phase of their stellar evolution \citep{Yajima13b}.
If so, the contribution of $\lya$ photons from the ISM to the global 21 cm signal can be negligible. Given the uncertainties we  
therefore also investigate the 21 cm signal without $\lya$ photon contribution from the ISM. 

The mean intensity is then estimated by
\begin{equation}
\Ja = \frac{\epsilon_{\lya} \times {\rm c}}{4 \pi}.
\end{equation}

\subsection{Spin temperature}

Finally we derive the 21 cm flux by considering the ionization, temperature and $\lya$ radiation field estimated in above equations. 
The differential brightness temperature to CMB is estimated by \citep{Furlanetto06a},
\begin{equation}
\delta T_{\rm b} = 28.1 \; {\rm mK} \; x_{\rm HI} (1+\delta) \left( \frac{1+z}{10} \right)^{\frac{1}{2}} \frac{T_{\rm s} - T_{\rm CMB}}{T_{\rm s}}
\label{eq:tb}
\end{equation}
where $\delta$ is the overdensity, $\Ts$ and $T_{\rm CMB}$ are the spin temperature and CMB temperature respectively.
Since we focus on the spatial mean of the 21 cm emission, we assume $\delta = 0$.
The spin temperature is changed by 
\begin{equation}
T_{\rm s}^{-1} = \frac{T_{\rm CMB}^{-1} + x_{\rm \alpha}^{-1}T_{\rm c}^{-1} + x_{\rm C}^{-1}T_{\rm gas}^{-1}}{1 + x_{\alpha} + x_{\rm C}}
\label{eq:ts}
\end{equation}
where $T_{\rm C}$ is the color temperature of the $\lya$ line, $\Tgas$  is the kinetic temperature of the gas, $x_{\rm \alpha}$ and $x_{\rm C}$  are the coupling coefficients of $\lya$ photon scattering and gas collision, respectively. These coefficients are estimated as \citep{Hirata06}
\begin{eqnarray}
\xa &=& 1.81 \times 10^{11} (1+z)^{-1} S_{\rm \alpha} \Ja, \\
x_{\rm C} &  = & \frac{T_{\star}}{A_{10} \Tcmb}(C_{\rm H} + C_{\rm p} + C_{\rm e}),
\end{eqnarray}
where  $S_{\rm \alpha}$ is the  scattering amplitude factor, $T_{\star} = 0.068~\rm K$, $A_{10} = 2.85\times 10^{-15}~\rm s^{-1}$ is the spontaneous emission factor of the 21 cm transition, $P_{\alpha}$ is the number of scatterings of $\lya$ photons per atom per second, $C_{\rm H}$,  $C_{\rm p}$ and $C_{\rm e}$ are the de-excitation rates due to collision with neutral atoms, protons, and electrons, respectively. We use the fitting formula given by \citet{Hirata06} for the $S_{\rm \alpha}$ and that by \citet{Liszt01} and \citet{Kuhlen06} for the de-excitation rates. 

Since $T_{\rm C}$ quickly settles into $\Tgas$ owing to the recoil effect of $\lya$ photon scattering, $T_{\rm C} = \Tgas$ is assumed in our calculations. The spin temperature depends sensitively on the coupling due to $\lya$ scattering and collision. When the coupling is strong, $\Ts$ becomes $\sim \Tgas$.  

\begin{table}
\begin{center}
\begin{tabular}{ccccccc}
\hline
  & $ a$ & $b$ & $c$ & $d$ & $\fesc$  \\
\hline
PopIIa     & 0.017   &0.13& 3.3 & 5.3 & 0.2  \\
\hline
PopIIb     & 0.017  &0.13& 3.3 & 7.0 & 0.2  \\
\hline
PopIIc     & 0.017   &0.13& 4.5 & 5.3 & 0.2  \\
\hline
PopIId     & 0.017   &0.13& 4.5 & 5.3 & 0.5  \\
\hline
PopIIe     & 1.6   &$7.0\times10^{-3}$& --- & --- & 0.2  \\
\hline
\end{tabular}
\caption{
Model parameters for the  Pop II SFRD.
For PopIIa-c, the parameters $a - d$ relate to the SFRD as a function of redshift via  ${\rm SFRD} = (a + bz)h / \left[ 1 + (z/c)^{d} \right]$.
The parameters of PopIIe are used for the exponential decay model at $z > 8$ using ${\rm SFRD} = b \times {\rm exp}(-az)$.
At $z \le 8$, the SFRD of the PopIIe is the same as for model PopIIa.
The parameter $\fesc$ is the escape fraction of ionizing photons from haloes. 
}
\label{table:model}
\end{center}
\end{table}

\begin{table}
\begin{center}
\begin{tabular}{ccccccc}
\hline
  & $A$ & $B$ & $\fesc$  \\
\hline
PopIIIa   & $5\times10^{-4}$ & 10.0  & 0.5  \\
\hline
PopIIIb   & $4\times10^{-3}$ & 1.0 & 0.5  \\
\hline
PopIIIc   & $5\times10^{-5}$ & 10.0 &  0.5  \\
\hline
\end{tabular}
\caption{
Model parameters for the Pop III SFRD.
The SFRD is modeled as ${\rm SFRD} = A \times{\rm exp} (- |z - 15|/ B)$.
}
\label{table:popiiimodel}
\end{center}
\end{table}

\begin{table}
\begin{center}
\begin{tabular}{ccccccc}
\hline
  & $ \tau_{\rm e}$ & $z_{\rm tr}$  \\
\hline
PopIIa + PopIIIa     & 0.087   &11.5  \\
\hline
PopIIa + PopIIIb    & 0.095  &15.2  \\
\hline
PopIIa + PopIIIc     & 0.052   &7.7  \\
\hline
PopIIa     & 0.048   & 7.5  \\
\hline
PopIId   & 0.086  & 10.4   \\
\hline
PopIIe+PopIIIa   & 0.084  & 11.0   \\
\hline
PopIIe   & 0.045  & 7.2   \\
\hline
\end{tabular}
\caption{
Thomson scattering optical depths of different models. $z_{\rm tr}$ is redshift at when the IGM temperature exceeds the CMB.}
\label{table:tau}
\end{center}
\end{table}

%
%

\section{Results}
\label{sec:result}

\subsection{Ionization history}

\begin{figure}
\begin{center}
\includegraphics[scale=0.45]{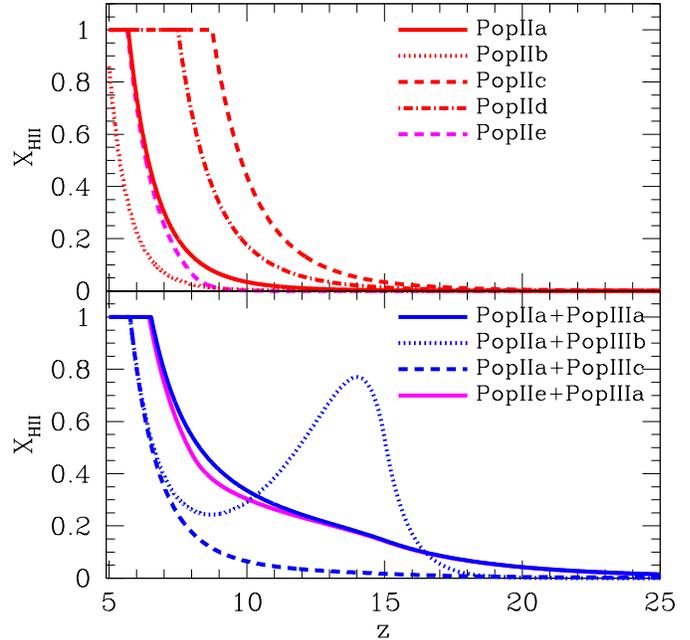}
\caption{
Redshift evolution of the ionization degree of the IGM with different models.
}
\label{fig:xh2}
\end{center}
\end{figure}

Ionization histories of the IGM are shown in Figure~\ref{fig:xh2}.
At $z > 15$, most of the IGM in  a neutral state for all models. 
At  $z < 15$, the ionization history of the individual models starts deviating from each other depending on their star formation histories. 
In the cases of Pop II stars alone, the ionization degree monotonically increases with time. 
For example, the case of PopIId reaches $\xhii = 0.5$ at $z = 9.8$,
while it is at $z=6.4$ in the case of PopIIa, because the ionizing photon emissivity of PopIId is higher than that of PopIIa by factor $\sim 5$. 
The ionization degree in the PopIIe model is very small at $z < 8$, then it catches up with that of PopIIa at $z \le 8$. 
Inclusion of Pop III stars in general changes the ionization degree to higher values as they provide an additional source of ionizing photons, and  shifts the redshift of reionization to higher $z$.

The PopIIa+PopIIIa model shows the IGM is started to be ionized earlier ($\xhii \sim 0.1$ at $z = 16.4$), 
then the ionization degree slowly evolves and reaches to $\xhii = 0.5$ at $z=8.3$.
In addition, the PopIIe+PopIIIa model also shows a similar ionization history. 
This is because  the ionizing photon emissivity of  Pop III stars is dominant at $z > 8.3$,
and that of the Pop II stars becomes almost the same as that in the PopIIa model. 
In the case of PopIIa+PopIIIb, the ionization history shows the more complex evolution. 
Due to the strong peak of the SFR of PopIII stars, at $z \sim 14$ most of gas is highly ionized once. 
As the PopIII SFR  decreases, the ionization fraction $\xhii$ drops back to a moderately neutral state $\xhii \sim 0.24$ at $z=8.6$.
At $z < 8$, the IGM is highly ionized again due to the PopII stars ($\xhii \sim 0.5$ at $z = 6.6$). 
In addition, despite the redshift evolution of the PopIII SFR  is symmetric with a maximum at  $z=15$, 
the redshift evolution of $\xhii$ is asymmetric. This is because of contribution by Pop II stars and the lower recombination rate at lower redshift which is $\propto (1+z)^6$.
In the case of PopIIa+PopIIIc, the contribution by PopIII stars is not significant. 
The ionization is somewhat enhanced by PopIII stars at $z \gtrsim 10$.

As the IGM is ionized, the UV background forms and can suppress star formation \citep{Susa08, Faucher10, Yajima12h}. 
Model  PopIId indicates that gas in galaxies is efficiently converted into stars, leading to a stronger UV radiation field. 
Recently \citet{Hasegawa13} showed that the SFRD is self-regulated by external and internal UV feedback in cosmological radiation-hydrodynamics simulations. 
Their simulated SFRD is similar to our model PopIIa. 
In addition, \citet{Johnson13} presented a co-evolution model of the SFRD for Pop II and III stars within a cosmological simulations including metal enrichment and radiative feedback. The SFRD of Pop III stars in their calculations is rather flat as a function of redshift which is similar to our models PopIIIa and PopIIIc. Their results show a  Pop II SFRD  that steeply increases with decreasing redshift which is similar to our PopIIa and PopIIc models.

\subsection{Thomson scattering optical depth}

\begin{figure}
\begin{center}
\includegraphics[scale=0.5]{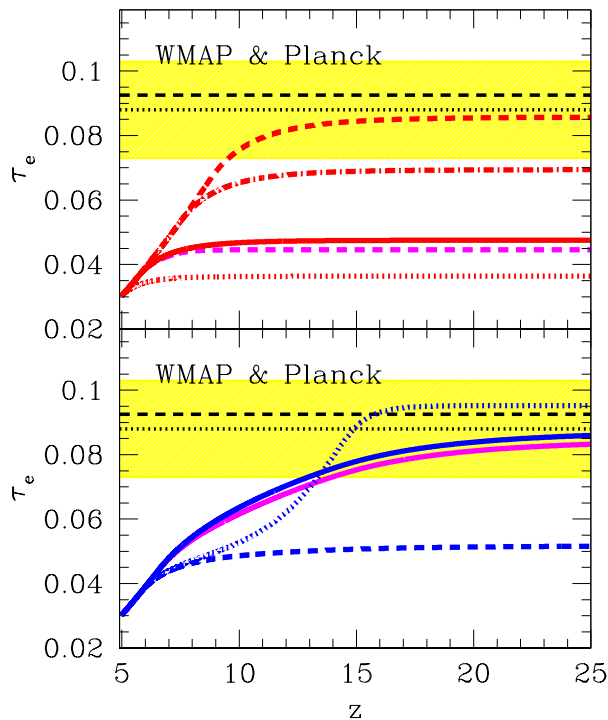}
\caption{
Thomson scattering optical depth for the same models shown in Figure \ref{fig:xh2}. 
Black dashed and doted lines are the observational data by WMAP (Komatsu et al. 2011) and Planck (Planck collaboration 2013) 
with the error of WMAP shown in yellow shade.
}
\label{fig:tau}
\end{center}
\end{figure}

The ionization histories in terms of the Thomson scattering optical depth ($\taue$) 
for the different models is shown in Fig.~\ref{fig:tau}. A subset of the models, most notably models with high PopII or PopII+PopIII star formation at $z > 10$, are in good agreement with the observational limits by the CMB  \citep{Komatsu11, Planck13}.  Inclusion of PopIII stars proves to be important due to their top heavy IMF and the associated higher contribution of ionizing photons per stellar mass formed. Increasing the PopIII star formation rate by one order of magnitude at all redshift increases the optical depth in model PopIIa+PopIIIc  from  $\taue= 0.052$ to $0.087$. The optical depth from PopII stars only in this model is $\taue = 0.048$.
 The upper limit PopII-only model PopIId is agreeing as well with the observational limits on $\taue$ owing to its high star formation rate.  In principle the uncertainty in the star formation history at $ z \geq 9$  allows for some of these models to boost the ionization fraction at $ z \geq 9$ by increasing the star formation history. Our results suggest, that the lower limit PopII star formation rate model PopIIa is not able to catch up with the CMB limits even if it would reach star formation rates similar to those in the upper limit model PopIId at  $ z \geq 9$. The fact that the optical depth is an integrated quantity over redshift results in a certain degree of smearing out of the different ionization histories. This is most apparent for the models  PopIIa+PopIIIa and PopIIa+PopIIIb which have  $ \taue=0.087$ and $0.095$, respectively.  The latter degeneracy makes the optical depth only a limited probe of the ionization history of the universe. As we will show in section \ref{sec:21cm} this degeneracy can be broken using 21 cm tomography.

In this work, we fix  $\fesc$ based on recent simulations, and investigate the 21 cm signal for different SFRD models. 
However,  $\fesc$ is highly uncertain due to the small sample of observed galaxies from which ionizing flux is detected \citep{Iwata09}.
Previous work investigating $\fesc$ for fixed SFRD from simulations or observations \citep[e.g.,][]{Wyithe10, Paardekooper13, Jaacks13}, 
finds $\fesc$ of $\sim 0.05 - 0.5$ depending on the model. 
The ionization history in degenerate in SFRD and $\fesc$ as   $\nion \propto {\rm SFRD} \times \fesc$. A possible way to break this degeneracy is by combining the ionization history with $\tb$ in future observations to get an accurate estimate of $\fesc$.

\subsection{Spin temperature}

\begin{figure}
\begin{center}
\includegraphics[scale=0.43]{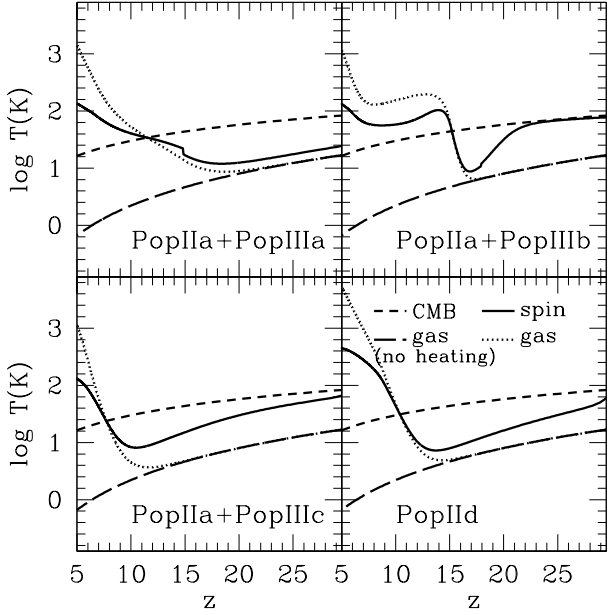}
\caption{
Redshift evolution of spin temperature with different models.
Dot and dash lines are gas and CMB temperature, i.e., $2.7\times(1+z)$ K, respectively.
Long dash lines are gas temperature without heating by stellar sources, 
i.e., only expansion cooling is considered in Equation~\ref{eq:temp}.
}
\label{fig:ts}
\end{center}
\end{figure}

Figure~\ref{fig:ts} shows the redshift evolution of CMB, gas and spin temperature in different models. 
The CMB temperature shows the expected cosmological $\propto (1+z)$ evolution, while the gas adiabatically cools via  $\propto (1+z)^{2}$. Once X-ray heating from supernovae is included in our models, the gas temperature gradually increases with increasing cosmic star formation rate exceeding the CMB temperature at lower redshifts. The gas temperature is going in tandem with the ionization history due to their mutual dependence on massive stars either providing ionizing photons or type II supernovae. Similar as in the case of the ionization history discussed in the previous section the heating rate depends on the amount of PopII and PopIII star formation, with the models  PopIIa+PopIIIa and PopIIa+PopIIIb exceeding the CMB temperature at $z \gtrsim 11$. The impact of Pop III star formation is highlighted again in the model PopIIa+PopIIIc, in which gas catches up to the CMB temperature only at $z=7.7$. This is close to the expected redshift of $z=7.5$ without any Pop III contribution in this model. In contrast the upper-limit Pop II star formation model predicts $z=10.4$ as the redshift at which the excess starts.  

As the gas is heated by X-ray photons, the spin temperature $\Ts$ starts deviating as well from  $\Tcmb$
 because  both, X-ray and $\lya$ photons are proportional to the SFR. 
It stays close to the gas temperature  $\Tgas$ because of a high $\lya$ background. 
At $z \gtrsim 15$ this is due to recombinations in  H{\sc ii} regions generated by Pop III stars.  Once Pop II star formation dominates at $z \lesssim 15$, the main source becomes continuum photons from stellar radiation. 
The small bump of $\Ts$ at $z\sim15$ in model PopIIa+PopIIIa is due to our assumptions about the emission of $\lya$ photons from the ISM which is turned off at $z\sim15$ as will be explained further in what follows. 
The coupling between  $\Ts$ and $\Tgas$  depends on the intensity of the $\lya$ radiation field as e.g. seen in model PopIIa+PopIIIc  where it is weaker at $z > 10$.  

\begin{figure}
\begin{center}
\includegraphics[scale=0.43]{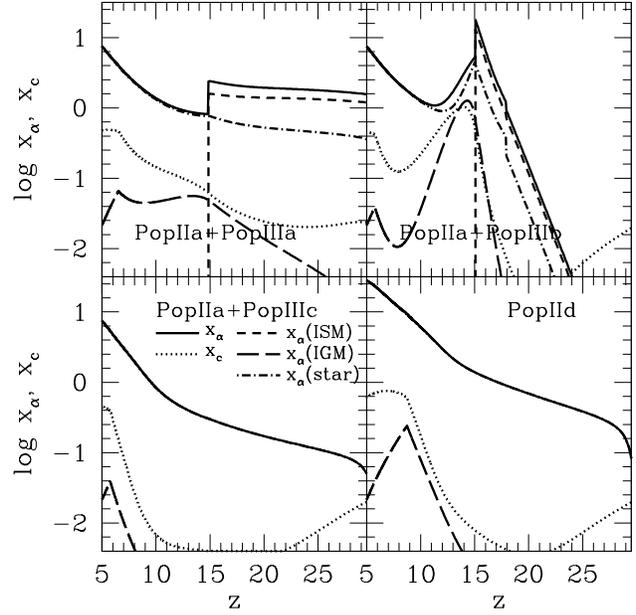}
\caption{
Redshift evolution of coupling coefficients. Solid and doted lines are by $\lya$ scattering and gas collision respectively.
 Dashed, long-dashed and dot-dashed lines are the coupling coefficients for $\lya$ photons considering 
 recombination in the ISM or IGM, or stellar continuum radiation alone, respectively. 
}
\label{fig:xa}
\end{center}
\end{figure}

We present in Figure~\ref{fig:xa} the detailed redshift evolution of the coupling coefficients for $\lya$ scattering $\xa$ and gas collisions $x_{\rm C}$. As shown in Equation~\ref{eq:ts} for values larger than unity for either coupling coefficients, the spin temperature is close to the gas temperature  resulting in either positive or negative 21 cm signals.  The coupling constant for collisions  $x_{\rm C}$ is always less than unity independent of the star formation model as it is driven by the density of the IGM, which is too low at the shown redshifts. Gas collisions thus play no important role in shaping the 21 cm signal at these redshfits. 
$\xa$ on the other hand is larger than unity at low redshifts for most of the models suggesting a strong coupling between $\Ts$ and $\Tgas$. Depending on the star formation model the contributions to $\xa$ are either dominated by recombinations in the ISM or continuum radiation from stars. 
The recombination signal from the IGM is always below that from the ISM and plays a minor role. The dominance of continuum radiation over ISM recombinations is closely related to the dominance of Pop II star formation over Pop III star formation because of the SED shape of Pop II stars which peaks at wavelengths beyond the Lyman limit. 
As seen in the PopIIa+PopIIIa and PopIIa+PopIIIb models, $\xa$ sharply drops down at $z \sim 15$ because of the way we model $\lya$ photons from the ISM. 
For model PopIIa+PopIIIa, the SFRD of Pop II stars stays above that of Pop III stars at $z \sim 15$, while $X_{\rm HII}$ becomes larger than 0.5 at $z \sim 15$ in the case of model PopIIa+PopIIIb. Therefore, the contribution from $\lya$ photons from ISM is turned off. This artificial model also affects the redshift evolution of the 21 cm signal, but it is negligible as shown in the next section.

\subsection{The 21 cm emission}
\label{sec:21cm}

\begin{figure*}
\begin{center}
\includegraphics[scale=0.6]{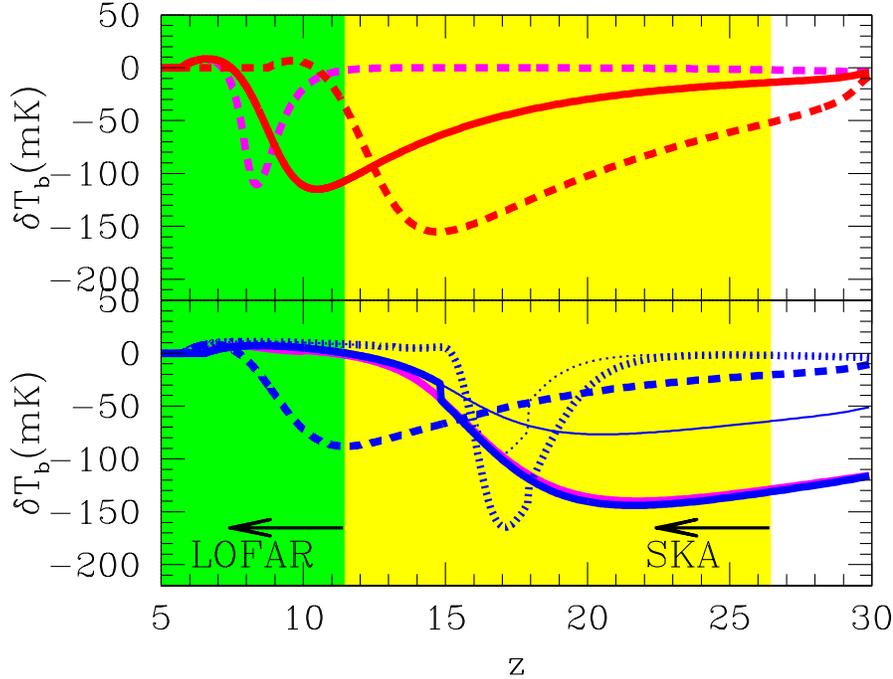}
\caption{
Redshift evolution of differential brightness temperature $\Tb$ for different models 
as shown in Figure~\ref{fig:xh2}.
Blue thin solid and dot lines represents the $\Tb$ without $\lya$ photons from ISM.
Shaded regions show the  range covered by LOFAR ($z \lesssim 11.4$) and SKA ($z \lesssim 26.4$).
}
\label{fig:tb}
\end{center}
\end{figure*}

Using the result from the previous sections we here derive the differential brightness temperature of the 21 cm line as a function of redshift.
Figure~\ref{fig:tb} shows the differential brightness temperature $\tb$ for different models. 
The colored regions show the ranges covered by  LOFAR and SKA. 
 Models PopIIa and PopIIa+PopIIIc fail to reproduce the observed value of $\taue \sim 0.048 - 0.052$ and have $\tb < 0$ even at $z \lesssim 10$. 
Our models that match the observed limits on $\taue$ show $\tb > 0 $ at $z \lesssim 10$, which is due to $\Ts$ and $\Tgas$ being higher than the CMB temperature.
Note however, $\tb \sim 0$ for model PopIId at $ z \lesssim 9$ because of the high ionization degree of the IGM. In contrast models PopIIa+PopIIIa and PopIIa+PopIIIb show $\tb > 1$ down to $z \sim 6-7$ because parts of the  IGM stay neutral, which matches the expectations from recent observations of LAEs that suggest some fraction of the IGM is still neutral at $z \sim 6.6$ \citep{Kashikawa11}.
However, a very small fraction of neutral hydrogen distributed around LAEs may change the $\lya$ flux and profile, leading to an underestimation of the ionization degree of IGM. 
The $\Tb$ of PopIIe+PopIIIa is very similar to that of PopIIa+PopIIIa. Hence, it suggests that the different SFRD of PopII stars at $z > 8$ cannot be resolved by the 21 cm signal for the case that Pop III stars are dominant ionizing sources.
The bump of $\Tb$ in model PopIIa+PopIIIa at $z\sim15$ is a result of the ceasing  $\lya$ photons from the ISM
which makes the coupling of $\Ts$ with $\Tgas$ weaker. 
In addition, we calculate  $\Tb$ without $\lya$ photon contribution from the ISM for models PopIIa+PopIIIa and PopIIa+PopIIIb
in which the $\lya$ photons from the ISM control  $\Ts$ significantly at $z \gtrsim 15$ as shown in Figure~\ref{fig:xa}. 
As a result, at $z \gtrsim 15$ the coupling of $\Ts$ with $\Tgas$ becomes somewhat weaker, 
and the absolute values of $\Tb$ decrease by factor $\sim 2$. However, the trends for the  redshift evolution of $\Tb$ does not change. 

The redshift evolution of our model PopIId is similar to that in previous studies\citep{Furlanetto06a, Mirocha13}, 
i.e., $\tb \sim 0$ at $z \gtrsim 20$, then negative due to the coupling of $\Ts$ and the low gas temperature with respect to the CMB, 
positive at $z \sim 10$ due to the heating, and $\tb \sim 0$ at $z \sim 6$ due to complete reionization of the IGM.
This evolution is clearly distinct to the one in our models including Pop III stars. As shown in Table~\ref{table:tau}, the gas temperature in the models with Pop III stars exceeds the CMB at higher redshift due to the efficient heating. 
Therefore,  models with efficient Pop III star formation, PopIIa+PopIIIa or PopIIa+PopIIIb, show earlier transitions of $\tb$ from negative to positive compared
to the models PopIIa+PopIIIc and PopIId.
The models with high Pop III SFR show that $\tb \gtrsim -50 ~\rm mK$ at $z \sim 15$, 
while Pop II stars alone does deep absorption $\tb \sim -170~\rm mK$.
In addition, if there is a strong peak in the  Pop III star formation and the IGM is reionized twice \citep{Cen03}, the transition of $\tb$ is sharp, which can be used as a signal for a burst in Pop III star formation. 

Future or ongoing 21 cm observation will give us information about the redshift evolution of $\tb$. 
LOFAR will cover the frequency range of $115-180~\rm MHz$ for the Epoch of Reionization project\footnote{ http://www.lofar.org/astronomy/eor-ksp/lofar-eor-project/lofar-epoch-reionization-project},
 which will be corresponding to the 21 cm signal from the IGM at $6.9 \lesssim z \lesssim 11.4$. 
In our models, Pop II stars are the dominant sources of  ionization and heating at these redshifts. 
Hence it is  difficult to disentangle the SFH of Pop III stars from  LOFAR data alone.
On the other hand, 
SKA will cover the frequency range of $50 - 300~\rm MHz$ ($3.7 \lesssim  z \lesssim 26.4$) 
\footnote{https://www.skatelescope.org/technology/} 
and  be able to distinguish different models, e.g., the late transition in PopIId and the sharp transition in PopIIa+PopIIIb.
Thus  measurements of $\tb$ via  the 21 cm signal  probe the SFH of Pop III and Pop II stars at high redshift.

%
%

\section{Discussion} 
\label{sec:discussion}

\subsection{X-ray heating from black holes}
In this work we have so far focused only on the X-ray contribution from SNe. Accreting black holes (BHs) are another viable source of X-rays that can modify the 21 cm signal \citep[e.g.,][]{Baek10, Mirabel11, Jeon14}.    
We approximate the contribution from black holes assuming they accrete close to the Eddington limit \citep{Willott10b}, and that the mass in BHs is  $\sim 10^{-3}$ times the stellar mass, which is the local observed value (e.g., \citealt{Marconi03, Rix04}). The bolometric luminosity of the BH is then  $L_{\rm Edd} = 1.4 \times 10^{35} {\rm erg \; s^{-1}} [M_{\rm star}/\Msun]$. 
The above luminosity is an upper limit given that feedback from the accreting black hole is able to reduce the accretion close to the black hole  \citep{Milosavljevic09a, Alvarez09, Li11, Park11}.
We take for the spectrum from the accretion disc around the black hole the following power-law  \citep{Laor93, Marconi04, Hopkins07},

\begin{equation}
\nu L_{\nu} \propto \begin{cases}
\nu^{1.2} & {\rm for} ~{\rm log}\; \left( \nu/{\rm Hz} \right) \le 15.2\\
\nu^{-1.2}
& {\rm for} ~{\rm log}\; \left( \nu/{\rm Hz} \right) > 15.2. 
\end{cases}
\end{equation}
Only X-ray photons with $E > 1~\rm KeV$ contribute towards the heating as softer X-ray photons mostly ionize hydrogen and helium  \citep{Furlanetto06a}. The former consist of  $1.3\times10^{-3}$ the total energy in X-ray photons. 

\begin{figure}
\begin{center}
\includegraphics[scale=0.43]{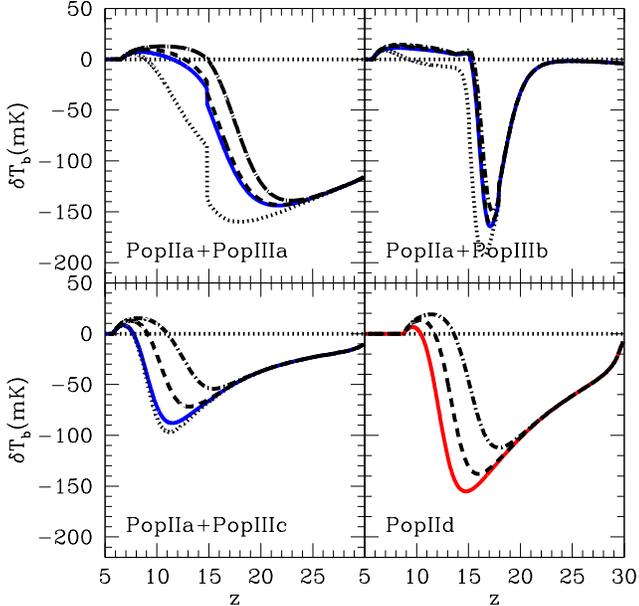}
\caption{
Redshift evolution of differential brightness temperature for different models.
Solid lines are the same as in Figure~\ref{fig:tb}.
Dotted lines show the case in which we assume that the Pop III and Pop II IMF both are of Salpeter shape and range between $0.1-60~\Msun$. 
Dashed and Dashed-dotted lines are the results for $\tb$ including X-ray heating from black holes with a mass ratio of $M_{\rm BH} / M_{\rm star}=10^{-3}$ and 
$M_{\rm BH} / M_{\rm star}=10^{-2}$, respectively.
}
\label{fig:tb_comp}
\end{center}
\end{figure}

The evolution for $\tb$ is shown in Figure~\ref{fig:tb_comp}.
Due to the heating by BHs, the gas temperature is higher, 
resulting in higher $\tb$. 
However, for models with efficient Pop III star formation (PopIIIa and PopIIIb), the addition of heating from BHs does not alter the results significantly as the heating from Pop III supernovae dominates. 
In contrast, models PopIIa+PopIIIc and PopIId show an increase in $\tb$ accompanied by the additional X-ray heating. 
For the model PopIIa+PopIIIc the lowest temperature changes only slightly from $\tb = -112.0$ to $-99.3~\rm mK$ and 
the transition redshift from negative to positive $\tb$ changes from $z=7.7$ to $9.1$.
Even for the model with the highest Pop II star formation rate (PopIId) the minimum temperature of $\tb$ changes from $-170.3$ to $-153.4~\rm mK$. Considering that we assumed accretion at the Eddington rate on to the BHs our results predict that BHs will not leave a significant imprint onto the 21 cm signal.

Recent observations indicate that the mass ratio of BHs to stellar mass in high-redshift galaxies might be larger than that of local galaxies \citep[see a review by][]{Kormendy13}.
Here we additionally investigate the case of a high ratio with $M_{\rm BH} / M_{\rm star} = 10^{-2}$ which is ten times larger than for local galaxies and predict the effects on the X-ray background.
The dashed-dotted lines shows  $\tb$ in this case.
Except for the PopIIa+PopIIIb, X-rays from BHs significantly change the redshift histories of $\tb$. 
In the case of PopIIa+PopIIIb, the IGM  quickly changes from cold and neutral with low stellar mass (low x-ray heating rate) to a hot ionized state due to the rapid Pop III star formation at $z \sim 15$. Hence, without the BH X-ray heating, the IGM is heated up to higher temperatures than that of the CMB. At $z < 15$, the X-rays from BHs heat the IGM further up. However, once the temperature becomes much higher than the CMB,  $\tb$ does not depend on the gas temperature significantly (see equation~\ref{eq:tb}). Hence the impact of BHs is smaller than in other models.
For model PopIIa+PopIIIa, the maximum $\tb$ changes from $8.0$ to $15.0$, and the minimum $\tb$ goes from $-151.4$ to $-147.0$. 
The transition redshift also changes from $11.5$ to $14.8$.
In addition, for model PopIId, the maximum $\tb$ changes from $6.8$ to $19.7$, the minimum $\tb$ goes from $-170.3$ to $-127.3$, 
and the transition redshift goes from $10.3$ to $13.5$.
As a result, the PopIId model roughly shows similar redshift evolution of $\tb$ as the models including Pop III stars due to the efficient X-ray heating. 
Therefore, in the case of very high BH mass to stellar mass  ratios and Eddington limited accretion onto the black hole,
it becomes difficult to distinct the SFRD of different populations due to the addition of another heating source.

\subsection{Pop III IMF} 

The SN heating rate for Pop III stars in our model is $\sim 17$ times higher than that for Pop II stars based on our choice of IMF slope and range. The exact form of the latter is still heavily debated \citep{Bromm09, Susa13, Stacy14}, and we estimate its impact in the limiting case of assuming both PopIII and PopII have the same IMF. 
 The doted lines in Figure~\ref{fig:tb_comp} show $\tb$ for models where the IMF of Pop III and Pop II stars are the same, i.e. models with lower heating rate from Pop III stars. Models PopIIa+PopIIIa and PopIIa+PopIIIb show the expected trend of negative 
$\tb$ down to lower redshifts compared to the fiducial models. The model with the gradual evolution of the Pop III star formation rate (PopIIa+PopIIIa) only heats the IGM to the CMB temperature by  $z=9$ in contrast to $z=11.5$ in the fiducial model. Please note that in the ``bursty'' Pop III star formation rate model (PopIIa+PopIIIb) $\tb$ is still negative even at the point when the Universe is reionized in this model at $z = 14.$ ($\xhii = 0.77$). The higher emissivity of ionizing photons by Pop III stars allows to ionize the IGM for lower star formation rates, respectively heating rates. The low Pop III star formation rate model (PopIIIc) only shows slight deviations due to the relatively small contribution from Pop III stars at any redshift. 

\begin{figure}
\begin{center}
\includegraphics[scale=0.4]{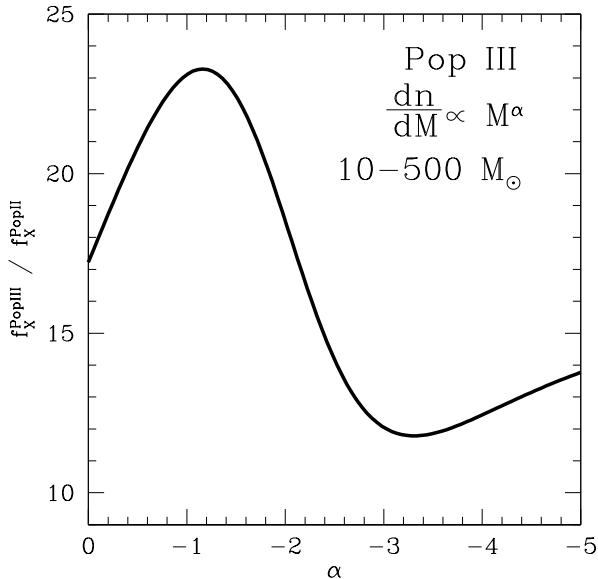}
\caption{
The ratio of  X-ray luminosity from Pop III and Pop II SN remnants  at the same SFR. 
The horizontal-axis shows the slope of the Pop III IMF,
i.e., $\frac{dn}{dM} \propto M^{\alpha}$.
The Pop III stars mass ranges from 10 to 500 $\Msun$, and the Pop II from 0.1 to 60 $\Msun$ and has slope $\alpha=-2.35$.
}
\label{fig:sn}
\end{center}
\end{figure}

As shown above, the heating rate by SNe changes with the shape of the IMF. 
Figure~\ref{fig:sn} shows the ratio of X-ray luminosity from Pop III SN remnants to Pop II SN remnants for varying Pop III  
IMF slopes, $\frac{dn}{dM} \propto M^{\alpha}$, in the mass range of $10-500~\Msun$.
It has a peak at $\alpha \sim -1.2$, and decreases with $\alpha$.
The X-ray luminosity of Pop III stars is determined by the mass fraction settling into supernova
and the number fraction between the core-collapse and pair-instability SN (Equation~\ref{eq:sn}).
At  $\alpha = -1.2$, 0.58 per cent of stellar mass ends in core-collapse SN, 
and 22.7 per cent in pair-instability SN, resulting in the net energy of $E_{\rm SN}=2.2\times10^{52}~\rm erg$ per supernova. 

\begin{figure}
\begin{center}
\includegraphics[scale=0.65]{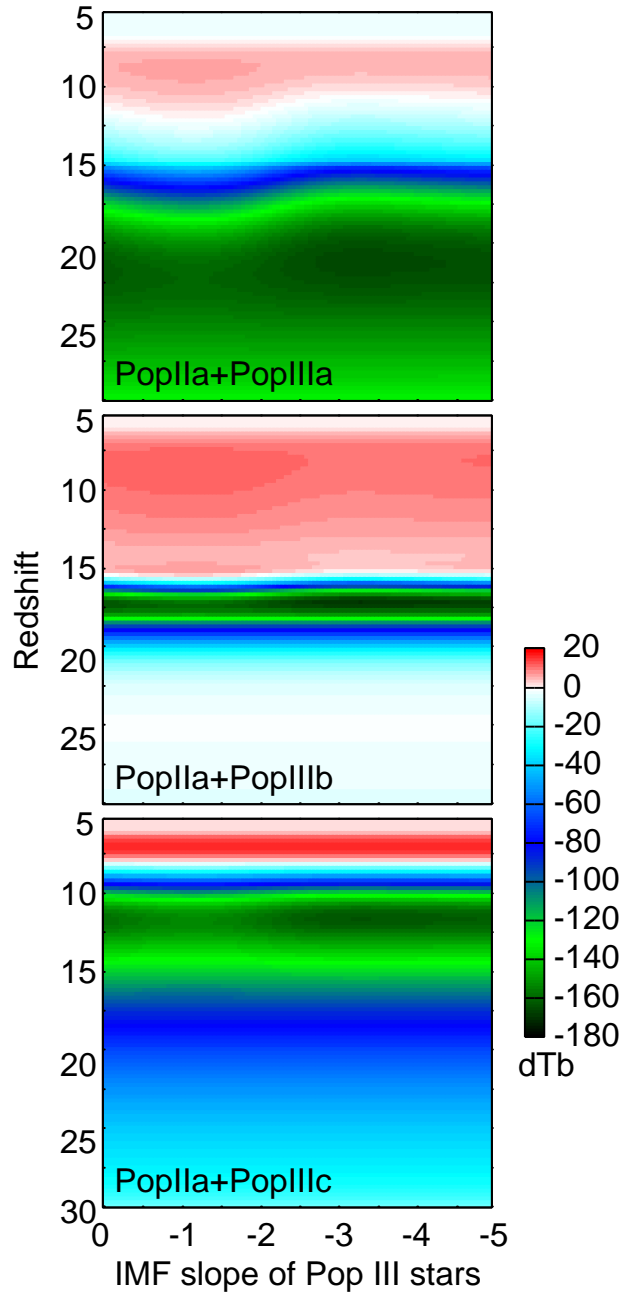}
\caption{
Two-dimensional map of the differential brightness temperature. 
The horizontal-axis is the slope of the initial mass function of Pop III stars,
the vertical-axis shows the redshift. 
}
\label{fig:2d}
\end{center}
\end{figure}

Figure~\ref{fig:2d} summarises the redshift evolution of $\tb$ for different IMF slopes. 
As can be seen in the figure, the effect is not very strong for the different models. Model PopIIa+PopIIIa shows a slight shift towards higher redshifts at which  $\tb \sim 0$ in models with shallow slope $\alpha \sim -1 $. Interestingly, for the model  PopIIa+PopIIIb the IMF slope has only very weak impact on $\tb $, even though the total mass in stars is comparable to the one in model PopIIa+PopIIIa. 
This is because partially ionized and heated gas phases are important for the 21 cm signal to distinguish the nature of the sources, 
however, the IGM  quickly changes from cold and neutral state  to a hot and  highly ionized state. This results in a small impact of the IMF.
Thus, it may be difficult to distinguish the IMF slope of Pop III stars by the 21 cm signal alone,
even if the Pop III SFR is high and their contribution to cosmic reionization is large. 
However, the insensitivity of the signal to the IMF allows to derive robust estimates for the Pop III SFH.

In addition, the IMF of Pop III stars can control the SFRD of Pop II and III via metal enrichment from stars and supernovae. 
A higher X-ray heating model with $\alpha \sim -1$ is associated with efficient metal enrichment, resulting in suppression of Pop III star formation \citep{Omukai05}. 
We will investigate the IMF, SFRD and the corresponding 21 cm signal in future work.

%
%

\section{Summary}
\label{sec:summary}
Considering the present observational limits on the star formation rate in the high redshift Universe, we generated models of Pop II and Pop III star formation that are both consistent with these limits and the Thomson scattering optical depth of the cosmic microwave background. To break the degeneracy between these models we predict the expected 21 cm signal due to the heating of the IGM from SN remnants. 

The limits on the SFR suggested by observed galaxies require high PopIII star formation rates to match the Thomson scattering optical depth. The cumulative  mass of Pop III stars in this case is $\sim 10^6 ~\rm \Msun~Mpc^{-3}$ at $z \sim 6$.  
Note that, in this work, we have assumed an escape fraction of $0.5$ for Pop III stars and $0.2$ or $0.5$ for Pop II stars based on 
recent numerical simulations \citep{Yoshida07, Wise09, Yajima11, Yajima12d, Paardekooper13}. These escape fractions can change the ionization history of the IGM and 
depend on halo mass \citep[e.g.,][]{Yajima11}. 
If we use a high escape fraction of 0.5 for Pop II stars and the SFRD derived from GRB observations which take low mass galaxies below the detection limit of galaxies into account, Pop II stars alone can cause cosmic reionization and match the recent WMAP or Planck data without the need for Pop III stars.

Due to the different IMFs of Pop III and Pop II stars the SN rate per solar mass of stars formed is higher in the case of Pop III
stars, as is the  $\lya$ photons production, resulting in a different thermal history of the 21 cm signal compared to cases with Pop II stars alone. 
At $z \lesssim 10$, it is difficult to distinguish different SFR models, 
because most of gas are already ionized and heated up to temperatures higher than the CMB. 
On the other hand, at $z \gtrsim 10$, depending on the contribution of Pop III stars to cosmic reionization, 
the 21 cm signal shows very different histories. 
The models with high Pop III SFR show that $\tb \gtrsim -50 ~\rm mK$ at $z \sim 15$, 
while Pop II stars alone does deep absorption $\tb \sim -170~\rm mK$.
Upcoming missions such as the Square Kilometre Array (SKA) will cover redshifts up to $z \sim 20$, 
and hence are able to give constraints on the SFR of Pop III stars. 

As well as SNe feedback, X-rays from BHs in galaxies can be heating sources of the IGM. 
We have studied the effect of BHs on the 21 cm signal with the assumption of 
the BH mass ratio to total stellar mass $M_{\rm BH}/M_{\rm star} = 10^{-3}$ and accretion at the Eddington rate. 
Our results show that  BHs could heat up the IGM, however the impact on the 21 cm signal was not so significant. 
In addition, we investigated the effect of different slopes for the  initial mass function  of Pop III stars.  
The fraction of mass ending up in core-collapse and pair-instability SNe changes with the IMF slope. 
As a result, depending on the IMF slope, the heating rate by Pop III SNe changes. 
The heating rate has a peak at an IMF slope $\alpha \sim -1.2$. 
Hence, for the models with high Pop III SFR, 
the transition redshift that $\tb$ goes from negative to positive, becomes earlier at $\alpha \sim -1.2$. 
However, the effect of different $\alpha$ is not very strong making it difficult to infer the IMF slope by 21 cm observation even if the SFR of Pop III stars is high. 

Here, we focused on the redshift evolution of the global 21 cm signal, i.e., the spatially mean value. 
In general the 21 cm signal significantly changes with the distance from ionizing/heating sources. 
We will study the inhomogeous structure of the 21 cm signal by using detailed cosmological simulations in future works. 

%
%
\section*{Acknowledgments}
We are grateful to Y. Li for valuable discussion.
We thank the anonymous referee for useful comments.

%
%

\bibliographystyle{mn}


\label{lastpage}

\end{document}